# The terrestrial late veneer from core disruption of a lunar-sized impactor

H. Genda[1], R. Brasser[1†], and S.J. Mojzsis[2,3*†]

[1]Earth-Life Science Institute, Tokyo Institute of Technology,
2-12-1 Ookayama, Meguro-ku, Tokyo 152-8551, Japan.
[2]Department of Geological Sciences, University of Colorado,
UCB 399, 2200 Colorado Avenue, Boulder, CO 80309-0399, USA.
[3]Institute for Geological and Geochemical Research,
Research Center for Astronomy and Earth Sciences,
Hungarian Academy of Sciences, 45 Budaörsi Street, H-1112 Budapest, Hungary.
*Corresponding author: Stephen J. Mojzsis (mojzsis@colorado.edu)
[†]Collaborative for Research in Origins (CRiO), The John Templeton Foundation –
FfAME Origins Program.

**Highlights:**

- Earth's Late Veneer is explicable via a single lunar-mass impactor after core formation.
- Elongation, disruption and disintegration of metallic cores occur in oblique collisions.
- Oxidation of sheared metal cores can suspend highly siderophile elements in the mantle.

*This paper was accepted for publication in Earth and Planetary Science Letter.*

**Abstract**

Overabundances in highly siderophile elements (HSEs) of Earth's mantle can be explained by conveyance from a singular, immense ($D$~3000 km) "Late Veneer" impactor of chondritic composition, subsequent to lunar formation and terrestrial core-closure. Such rocky objects of approximately lunar mass (~0.01 $M_⊕$) ought to be differentiated, such that nearly all of their HSE payload is sequestered into iron cores. Here, we analyze the mechanical and chemical fate of the core of such a Late Veneer impactor, and trace how its HSEs are suspended – and thus pollute – the mantle. For the statistically most-likely oblique collision (~45°), the impactor's core elongates and thereafter disintegrates into a metallic hail of small particles (~10 m). Some strike the orbiting Moon as sesquinary impactors, but most re-accrete to Earth as secondaries with further fragmentation. We show that a single oblique impactor provides an adequate amount of HSEs to the primordial terrestrial silicate reservoirs via oxidation of (<m-sized) metal particles with a hydrous, pre-impact, early Hadean Earth.

# 1 Introduction

Highly siderophile elements (HSE), which include the platinum-group elements (Os, Ir, Ru, Rh, Pt, Pd), as well as Au, and Re, are depleted in the Earth's mantle relative to chondrites (e.g. Becker et al. 2006). This is the expected outcome because as metallic cores grow within large and differentiated planetary embryos, HSEs should be effectively stripped from silicate mantles via strong metal-silicate partitioning of the metal component into metal phases; these then become segregated into the growing metallic cores. Yet, oddly, the abundances of these metals are far greater than what is predicted by high-pressure and high-temperature experiments for partitioning of HSEs between liquid metal and silicate (e.g. Mann et al. 2012; Rubie et al., 2015a,b and references therein). Adding to this riddle, the relative abundances of HSEs in the terrestrial mantle – and that of Mars (Brasser and Mojzsis, 2017) – are nearly chondritic despite their significantly different silicate-metal partitioning coefficients.

Hence, the observed HSE signature in the Earth's mantle is in conflict with the widely-held expectation that these elements should have been removed from the mantle during core formation (Kimura et al. 1974). To account for this observation, it was already proposed long ago that a relatively small augmentation (~1 wt%) of material enriched in HSEs was supplied late to the Earth's mantle after core formation was complete, in the form of a "Late Veneer" or LV (Chou, 1978; Rubie et al. 2016; Frank et al. 2016 and references therein). This exogenous explanation for Earth's mantle HSE signature is, however, not unanimously accepted, and debates with alternative scenarios are on-going (Willbold et al. 2011; Touboul et al. 2012; Puchtel et al. 2014; Touboul et al. 2014; Willbold et al. 2015; cf. Righter et al. 2015).

Several theories have been advanced to explain the source of the putative LV impactor(s): chondritic material supplied from the asteroid belt (Drake and Righter, 2002), leftover planetesimals from terrestrial planet formation that originated near Earth's orbit (Raymond et al. 2013), and fragments dispersed onto heliocentric orbit during a phase of giant impacts that created Earth and Venus from Mars-sized protoplanets (Genda et al. 2017a). It is noteworthy that recent data are consistent with an inner solar system origin of the LV object(s) (Fischer-Gödde and Kleine, 2017). If the LV was accreted in chondritic proportions, the total supplementary mass to Earth can be estimated at 0.5-0.8 wt%, (Walker, 2009; Day et al. 2016). For clarity and consistency with previous work (Brasser et al., 2016; Brasser and Mojzsis, 2017), we refer to mass added to the inner planets after the Giant Impact that formed the Moon (ca. 4.51 Ga) and the intervening time assigned by us to the LV epoch (between ca. 4.51 Ga and ca. 4.4 Ga) with the generic expression "late accretion".

In stark contrast to Earth, HSEs in the lunar mantle appear to indicate that the Moon acquired only an additional 0.02–0.035 wt% (Day & Walker, 2015; Kruijer et al. 2015). Based on this observation, the ratio of the accreted mass between Earth and the Moon is tentatively taken to be 1950 ± 650, which is two orders of magnitude higher than the ratio of their gravitational cross sections. Several studies have attempted to explain this extraordinarily high ratio. Bottke et al. (2010) proposed that the size-frequency distribution of the remaining planetesimals from planet formation had to have been shallow even at larger diameters in excess of ~1000 km. They further proposed that most the mass delivered to the Earth should have come from a few large objects comparable in size to asteroid 1Ceres (~945 km diameter), or more statistically likely, from a larger solitary object. With only a few such entities dwelling in the inner Solar System at that time, the Moon can statistically avoid accretion of large objects while the Earth did not (Sleep et al., 1989).

The effect of such a shallow size-frequency distribution was investigated in more detail by Brasser et al. (2016). Through a combination of $N$-body and Monte Carlo simulations, they concluded that a single lunar-sized body striking the Earth between the final giant impact event that created our Moon at ca. 4.51 Ga (Barboni et al. 2017 and references therein), and the last major terrestrial differentiation event that separated the silicate reservoirs at ca. 4.45 Ga (e.g. Allegre et al. 2008), can account for the HSE excess in the Earth's mantle. More recently, Brasser and Mojzsis (2017) also show that it is statistically unlikely that the Moon and Mars experienced an impact as large as Earth's to explain their respective HSE supplies, which as previously mentioned comports with the very low lunar HSE abundance compared to Earth (Day and Walker, 2015) and the relatively modest martian abundances (Day et al., 2016).

Like the asteroid 4Vesta (525 km diameter) and some parent bodies of the iron meteorites, a lunar-size body is expected to be differentiated depending on its formation timescale (e.g. Lee & Halliday 1997) which means almost all its HSEs are already partitioned into an iron core. It makes sense to understand further how core materials in such objects behave under various impact conditions. How are HSEs, sequestered in an impactor's core, able to pollute the terrestrial mantle and stay suspended there (Maier et al., 2009; Frank et al., 2016) rather than immediately sink to Earth's core (Stevenson, 1981)? If the impactor's core simply merges with the Earth's core, the impactor is unlikely to supply any HSEs to the Earth's mantle since little or no interaction with silicate reservoirs (crust and/or mantle) would occur (Dahl & Stevenson, 2010).

Here, we report on our investigations of the fate of the iron core of a lunar-sized planetary embryo[1] striking the early Hadean (pre-4.4 Ga) Earth that can account for the Late Veneer hypothesis. We test the premise that the impactor's core is temporarily ejected and thereafter sheared and fragmented into small particles that descend as hail of molten iron in the post-impact phase. This scenario leaves open the possibility of supplying (and suspending) HSEs to Earth's mantle via oxidation reactions by a primordial surface hydrosphere (Abe, 1993) and/or a relatively high oxygen fugacity fayalite+magnetite+quartz (FMQ)-buffered early Hadean mantle (Trail et al., 2011).

Our study involved impact simulations as well as examination of the post-collision evolution of the impactor's core materials with a postulated early Hadean Earth's hydrosphere and mantle. In Section 2 we describe our numerical methods and explain the outcome of the collision simulations. In Section 3 we test the degree to which the impactor's core is expected to accrete onto the Moon with the aid of *N*-body simulations. A discussion in Section 4 presents our observations of the outcomes of our model runs and Section 5 reports our conclusions.

## 2 Impact Simulations

Our simulations considered a lunar-sized impactor that struck the Earth under different impact angles in the period of late accretion for which previous mass and velocity analyses were reported (Brasser et al., 2016). In our model, we assumed that both Earth and the impactor are differentiated so that almost all of their HSE complements are already partitioned into their respective iron cores. We discuss the fate of the impactor's disrupted metallic core during a subsequent re-impact stage onto the Earth.

### 2.1 Numerical Methods

To perform our impact simulations, we use the smoothed particle hydrodynamics (SPH) method (e.g. Monaghan, 1992), which is a flexible Lagrangian method of solving hydrodynamic equations. Our numerical code is the same as that reported in Genda et al. (2015a). Our code can calculate a purely hydrodynamic flow with no material strength. The shock-induced pressure in our numerical setting described below is above 200 GPa, which is much higher than the Hugoniot elastic limit for typical rocks and iron (< 10 GPa, e.g., Melosh 1989). Thus, the assumption with no material strength is valid.

The masses of the target (Earth) and impactor were set to be 1 $M_⊕$ (= 6.0 × 10$^{24}$ kg, the Earth mass) and 0.01 $M_⊕$, respectively. As mentioned above, both the target and projectile are differentiated in our models. We further assume both objects have a 30% iron core and 70% silicate mantle by mass. According to recent detailed isotopic analysis of the Earth and meteorites, the accreted materials during the very last stage of the Earth's formation (i.e., LV) would be dominantly composed of enstatite chondrites (Dauphas 2017). Owing to the fact that enstatite chondrites contain about 30 wt% iron in the reduced form (e.g., Wasson and Kallemeyn 1988), a 30% iron core is considered for an impactor in our simulations. For the equation of state (EOS), we used the Tillotson EOS (Tillotson, 1962), which has been widely applied in other previous studies including planet- and planetesimal-sized collisional simulations (e.g. Canup and Asphaug, 2001; Citron et al., 2015). We used the parameter sets of basalt for the mantle and iron for the core (Benz and Asphaug, 1999).

---

[1] We propose that the Late Veneer object be named *Moneta* after the Roman goddess of memory and protectress of funds (money). Alternatively, in the Greek pantheon her name is *Mnemosyne*, a Titaness and the daughter of *Uranus* and *Gaia*. The rationale behind this proposal is that the Late Veneer event should neither be conflated with the Moon-forming event (by an object dubbed *Theia*), nor with the purported "late heavy bombardment".

The von Neumann–Richtmyer-type artificial viscosity was introduced as a hydrodynamics solver to capture shock waves (Fukuzaki et al. 2010; Genda et al., 2015b; 2017b). Conventionally, all SPH particles in our simulations have the same mass and the total number of particles used for the target and impactor was fixed at $10^6$ and $10^4$, respectively. The mass of each SPH particle is $6.0 \times 10^{18}$ kg, which corresponds to 160 km and 100 km in diameter for solid (or liquid) basalt and iron, respectively. We used $1 \times 10^5$ J/kg for the initial specific internal energy for both impactor and target. We calculate all impact simulations for $10^5$ sec (~ 1 day), and it takes about two weeks to perform one impact simulation with 16 cores. We considered no spin for pre-impact objects. Because the surface velocity of the spinning impactor and target is much less than the impact velocity, we expect that results do not change very much by adopting this approach (e.g., Canup 2008). To elaborate: Based on the spin period (> 2 hours) of the asteroids with $D > 100$ km (Warner et al. 2009), the surface velocity of a 3000 km-sized impactor with 0.01 $M_\oplus$ is estimated to be < 1 km/s. When the Moon orbited at 25 Earth radii, the spin period of Earth is estimated to be ~ 8 hours (Touma and Wisdom 1994), which corresponds to ~ 1 km/s for the surface velocity of Earth. As such, both surface velocities for the impactor and Earth are much less than the impact velocity (16 km/s) considered here. Consequently, we consider that the practical effects of the spin differences are minor.

We undertook eight impact simulations for various impact angles $\theta = 0°$ (a head-on collision), 30°, 35°, 40°, 45°, 50°, 55° and 60°. The impact velocity ($v_{imp}$) was kept constant at 16 km/s, which is the mean value obtained from $N$-body simulations of leftover planetesimals from terrestrial planet formation (Brasser et al., 2016). The probability of impact at an angle between $\theta$ and $\theta + d\theta$ is expected to be ½ $\sin^2 \theta d\theta$ (Shoemaker 1962). Importantly for the outcomes of this study, the above probability function has a peak at $\theta = 45°$, such that half of the impact events happen in the tails of the probability distribution between $\theta = 30°$ and 60°.

## 2.2. Collision Outcomes

Snapshots from our SPH simulations for four types of collisions at different impact angles noted above are presented in **Figure 1**. In the $\theta = 0°$ case (head-on collision, see **Fig.1a**), a symmetric cratering process is evident and the impactor's core quickly (< 1 h) merges with the Earth's core; also shown in the simulation is that a very small number of iron SPH particles (20 particles) are stranded in the Earth's mantle.

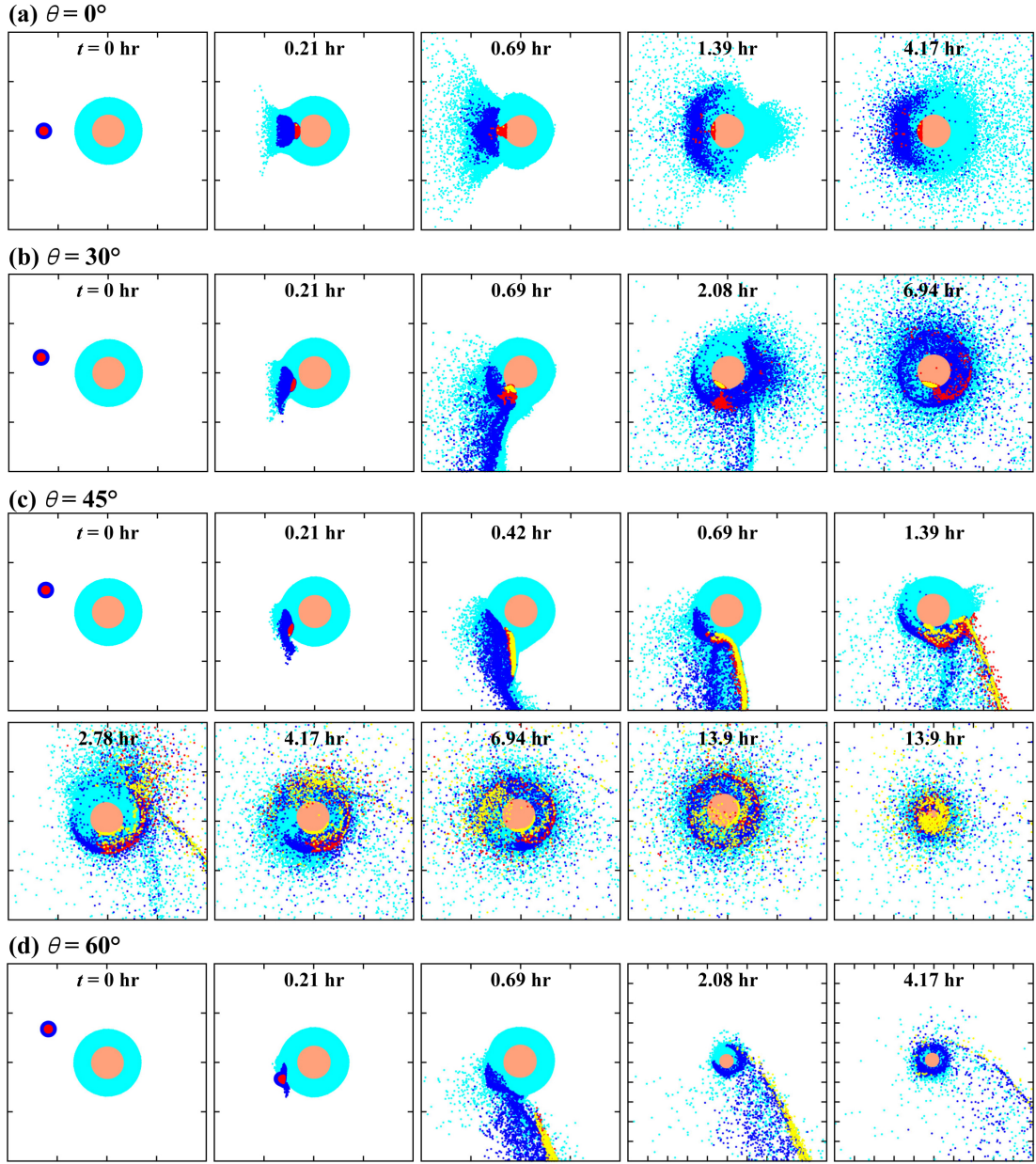

**Figure 1**. Snapshots for a collision of a large impactor of 1% $M_\oplus$ onto Earth. The impact velocity is 16 km/s. The impact angles are 0° (a), 30° (b), 45° (c), and 60° (d). Mantle and core materials for the impactor are colored by blue and red, respectively, and those for Earth are by light blue and orange. Impactor's core materials that have experienced fragmentation are colored yellow. The snapshots are not cross sections, but all SPH particles are layered on top of one another in order of Earth's mantle, impactor's mantle, Earth's core, impactor's core (red and then yellow), so that iron particles are clearly seen. The interval of tics in each snapshot is $1 \times 10^7$ m.

The amount of stranded iron materials would increase owing to the fact that the numerical resolution of each of our simulations is limited, and the standard SPH code has difficulty dealing with Rayleigh-Taylor and Kelvin-Helmholtz type instabilities (e.g. Hosono et al. 2016). Although the numerical simulation here is unable to provide the exact fraction of iron particles stranded in the target Earth's mantle, our work shows (as expected) that a large fraction of the impactor's core straightforwardly and quickly merges with the Earth's core in the head-on scenario.

In the first intermediate $\theta = 30°$ case (**Fig. 1b**), output shows that the behavior of the impactor's core is broadly like that of the head-on collision, but a higher fraction of the impactor's iron core (80 particles) is entrained in the Earth's mantle.

With regards to the statistically most-likely case (Shoemaker, 1962) of an impact angle of $\theta = 45°$ (**Fig. 1c**), our results show that the impactor undergoes elongation after the initial collision with the target Earth, and that because of this elongation a significant fraction of impactor material is ejected from the collision site. A sizeable quantity of ejected material thereafter re-accretes onto Earth in a metallic hail of secondaries (e.g. Melosh & Vickery, 1991; Johnson et al., 2014). The pressure of shocked materials of the impactor's core is released during their ejection, and therefore the iron blobs should fragment into smaller pieces due to shear velocity. In the next subsection, we estimate the sizes of suspended iron drops based on simple analytical considerations of this fragmentation.

In the second intermediate case of $\theta = 60°$ (**Fig. 1d**), the initial behavior resembles that for $\theta = 45°$. A large fraction of disrupted impactor flies off, however, so that most of the impactor's metallic payload that carries HSEs becomes lost. As a result, few of the impactor's iron SPH tracer particles remain to re-accrete.

**Figure 2** shows the mass fraction of impactor material gravitationally bound to Earth after collision for the various impact conditions explored above. The output of our simulations clearly demonstrate that steeper impact-angles of collision lead to larger fractions of impactor material being accreted. Significant escape of the impactor's rocky mantle occurs for $\theta > 30°$, but the escape of iron core material becomes important only for $\theta > 45°$. Since the impactor's iron core is initially enveloped by its rocky mantle, iron core material tends to be more difficult to escape from the system compared to silicate mantle.

Fittingly, we find that in the statistically most-likely case collision of a single lunar-size impactor at $\theta \sim 45°$ with Earth yields the most promising returns in terms of furnishing HSEs to the Earth's mantle in a Late Veneer (cf. Dahl and Stevenson, 2010). It has long been known (Gilbert, 1893) that an angle $\theta = 45°$ is the most likely impact trajectory, which lends some credence to our model for the origin of the Late Veneer from such a process. Next, we examine more closely the nature of the fragmentation regime of ejected iron material, and makes estimates of fragment size. Fragment size is an essential parameter when addressing the efficacy of oxidation of the re-accreted metal onto the crust and mantle in reaction with a postulated hydrous early Hadean Earth.

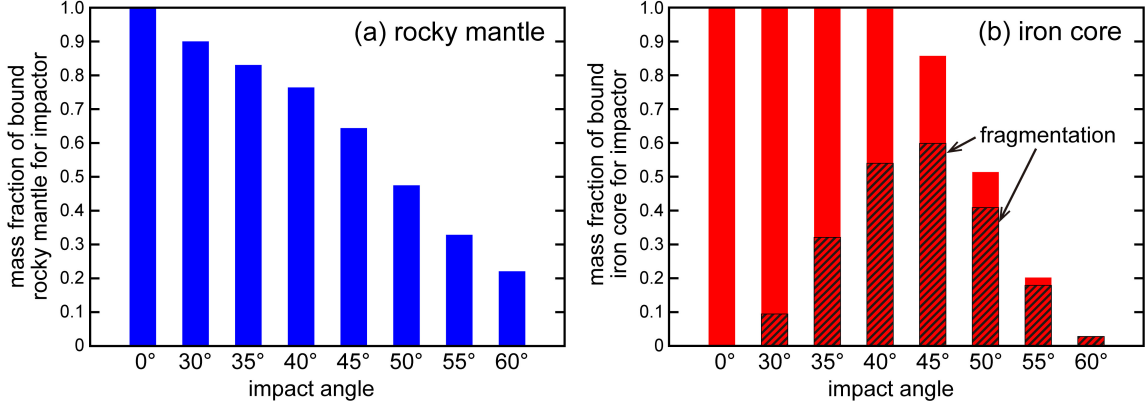

**Figure 2.** Mass fraction of impactor material that is gravitationally bound to Earth after collisions as a function impact angle. The left panel pertains to the impactor's rocky mantle material while the right panel is for the iron core. Some of the impactor's core materials that are gravitationally bounded to Earth experiences pressure release after the shock, which leads to its fragmentation (shaded region).

## 2.3 Post-collision Fragmentation of Impactor's Iron Core

For a common oblique collision ($\theta \sim 45°$; **Figure 1c**), the impactor's iron core experiences very high peak pressure (290 GPa on average) due to the passage of a shock wave that is generated during the impact. This high peak pressure is above the pressure for shock-induced incipient melting (~220 GPa) and complete melting (~260 GPa) for iron (Melosh 1989). Therefore, most of impactor's iron core is molten, and is subsequently pulled apart into an elongated shape and ejected from the Earth, which causes pressure release. During this process, fragmentation of impactor's iron materials commences.

Since the evolution of pressure for all SPH particles is embedded in our impact simulations, we can estimate the fraction of impactor's iron core that is fragmented. Fragmentation occurs when the density of the iron SPH particles drops below the normal density for the molten iron during adiabatic expansion with shear. During this expansion, the pressure also drops from shock-induced pressure (~ 300GPa) to extremely low pressure (~ 0GPa). We assume that fragmentation occurs when the pressure of a shocked SPH particle decreases down to 0.1 MPa (= 1 bar) during pressure release process. Using this criterion for the fragmentation, Figure 1 also shows the impactor's iron SPH particles (in yellow color) that experience the fragmentation. **Figure 2** also shows the fraction of the gravitationally-bound impactor's core that experiences fragmentation. The total mass of fragmented iron material peaks at $\theta = 45°$, such that approximately 75% of the re-accreted impactor's core is fragmented. On the other hand, at conditions where $\theta > 45°$ the mass of fragmented iron core *increases* with $\theta$, but most of them escape from Earth. As a result, total reaccreted mass of fragmented iron core *decreases* with $\theta$.

Due to the limit of the resolution for our numerical simulation, however, we are unable to directly calculate the fragmentation size of iron materials ejected by collisions, which is expected to be smaller than the size of a single SPH particle. Nevertheless, we can roughly estimate the typical size of fragmented and sheared iron blobs by considering the balance between surface tension of liquid iron ($\sigma$) and the local kinetic energy ($1/2mv^2$) induced by local shear velocity $v = \dot{\varepsilon}d$, where $\dot{\varepsilon}$ is the strain rate of the expanding melt, and $d$ is the typical size of molten droplets during fragmentation. According to Melosh and Vickery (1991), the typical size ($d$) of molten droplets ejected by an impact is given by:

$$d = \left(\frac{40\sigma}{\rho\dot{\varepsilon}^2}\right)^{1/3}. \qquad (1)$$

For liquid iron, $\sigma$ is ~ 2 N/m (Keene 1993), and the density of the iron droplets $\rho$ is ~ 7000 kg/m$^3$.

We calculate $\dot{\varepsilon}$ at the position $x$ through:

$$\dot{\varepsilon}^{\alpha\beta} = \frac{1}{2}\left(\frac{\partial v^\alpha}{\partial x^\beta} + \frac{\partial v^\beta}{\partial x^\alpha}\right), \qquad (2)$$

where the superscripts $\alpha$ and $\beta$ are the directions of the $x$-, $y$-, or $z$-axis. In the SPH method, the strain rate tensor of the $i$-th SPH particle can be calculated by summing up the contributions of the neighboring SPH particles (Benz and Asphaug 1994) such that:

$$\dot{\varepsilon}_i^{\alpha\beta} = \frac{1}{2\rho_i}\sum_j m_j\left[\left(v_j^\alpha - v_i^\alpha\right)\frac{\partial W_{ij}}{\partial x_i^\beta} + \left(v_j^\beta - v_i^\beta\right)\frac{\partial W_{ij}}{\partial x_i^\alpha}\right], \qquad (3)$$

where $m$ is the mass of the SPH particle, $\rho$ is the density, and $W$ is the kernel function (Monaghan and Lattanzio 1985). The subscript $j$ represents neighboring SPH particles, whose number was set to 64 in our calculations. We calculated the effective strain rate $\dot{\varepsilon}$ from the strain rate tensor $\dot{\varepsilon}^{\alpha\beta}$ as follows:

$$\dot{\varepsilon} = \left[\frac{1}{2}\sum_{\alpha\beta}\dot{\varepsilon}^{\alpha\beta}\right]^{1/2}. \qquad (4)$$

**Figure 3** shows the distribution of $\dot{\varepsilon}$ at the fragmentation regime calculated in our SPH simulations. The calculated strain rate is ~ 0.001–0.01 s$^{-1}$, which is consistent with the value estimated by the following simple physical consideration; $\dot{\varepsilon}$ ~ $v_{imp}/D_{imp}$ = 0.005 s$^{-1}$, where $v_{imp}$ = 16 km/s and $D_{imp}$ ~ 3000 km (the impactor's diameter in our simulation). Using the strain rate, we can estimate the size of fragments by Eq. (1). **Figure 3** also reports the size distribution of the fragment size of impactor's iron materials. We found that the typical size of iron fragments is ~10 m. Although some gravitational re-accumulation of iron fragments would happen during and/or just after the fragmentation, this effect would be minor. This is because 10m-sized iron fragments move apart at the velocity of ~ 0.05 m/s (=$\dot{\varepsilon}d$), which is larger than the two-body escape velocity of 10m-sized iron fragments (~ 0.01 m/s).

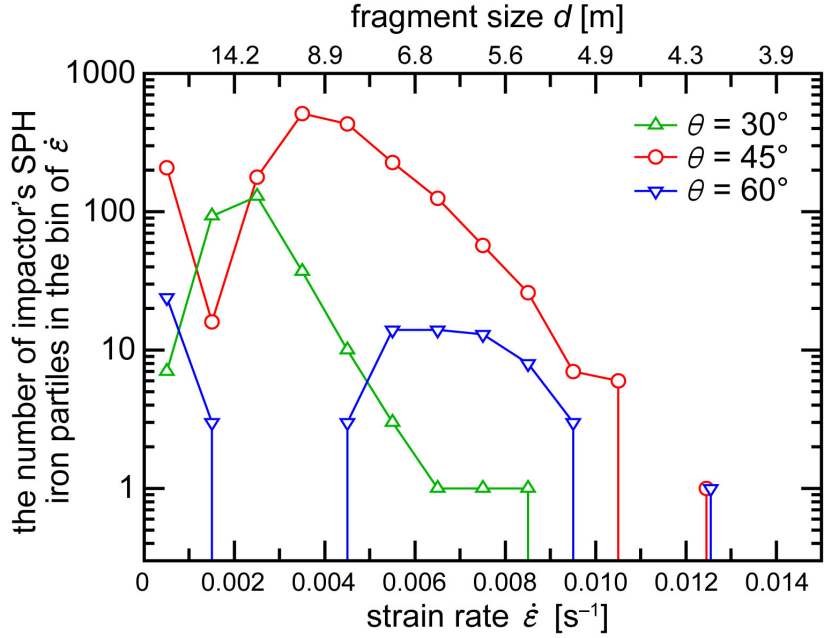

**Figure 3.** Distribution of the strain rate ($\dot{\varepsilon}$) for the impactor's core material at the time of fragmentation. The bin size of $\dot{\varepsilon}$ is 0.002. The fragment size is estimated by the balance between the surface tension and strain rate (see Eq.(1)).

In conclusion, an oblique impact of a lunar-sized differentiated planetary embryo at the typical impact angle of 45° can deliver approximately 75% of its iron core onto the near Earth's surface zone (which includes the lithosphere and upper mantle). We also find that this mass estimate is of the same order as the late accretion component on Earth computed from HSE concentrations and radiogenic W and Ru of the terrestrial mantle (Walker, 2009; Willbold et al., 2011; Fischer-Gödde and Kleine, 2017).

## 3 Re-accretion of Ejected Iron Materials onto the Moon

We now discuss the post-collision dynamical fate of the iron ejecta and evaluate the fraction that accretes onto the Moon as a "sesquinary" hail of molten metal. After the lunar-sized impactor (dubbed *Moneta*) struck the Earth, our simulations show that much terrestrial and impactor material is launched into Earth orbit, such that approximately half of the ejecta end up in hyperbolic orbits and will escape from Earth's gravity while the other half is bound on orbits with a period of just a few hours. Our analysis shows that there is a non-negligible fraction with a larger semi-major axis that could impact the Moon (cf. Gladman et al., 1995; Ramsey and Head, 2013).

To evaluate whether this sesquinary material could account for the lunar HSE inventory, we proceed as follows: The geocentric state vectors of iron core ejecta material in each of our simulations were saved once they were located at 2 Earth radii. We then merged together the iron particles from the SPH simulations with all impact angles, with most coming from the case with $\theta = 45°$. Not all of the ejecta material ends up at the 2 Earth radii cut-off distance at the same time, but for the sake of simplicity we shall assume that it is so. The state vectors were converted to geocentric orbital elements and the ejecta are treated as test particles. We placed the Moon on a circular orbit at either 25 or 35 Earth radii, which spans the approximate lunar semi-major axis some 50 Myr after lunar formation in the giant impact (e.g. Touma and Wisdom, 1994; Brasser et al., 2013). The inclination of the Moon ($i$) ranged from 0° to 180° at 30° intervals. The argument of perigee, longitude of the node and mean anomaly were uniformly random between 0° and 360°. We generated 1000 different initial orbits for the Moon per value of the inclination to properly account for a random orientation at the time of the impact. The simulations were performed with the SWIFT RMVS3 integrator (Levison and Duncan, 1994) with a time step of 2.62 minutes. The gravitational effects of the Sun and other planets were neglected in our simulations. Test particles were eliminated when they were farther than from the Earth than its Hill radius, or impacted either the Earth or Moon. The effects of the Earth's flattening were not included since most test particles had orbital periods of less than a day. The simulation time was set to 1 year to properly account for the few particles found to have relatively long-period orbits of a few months.

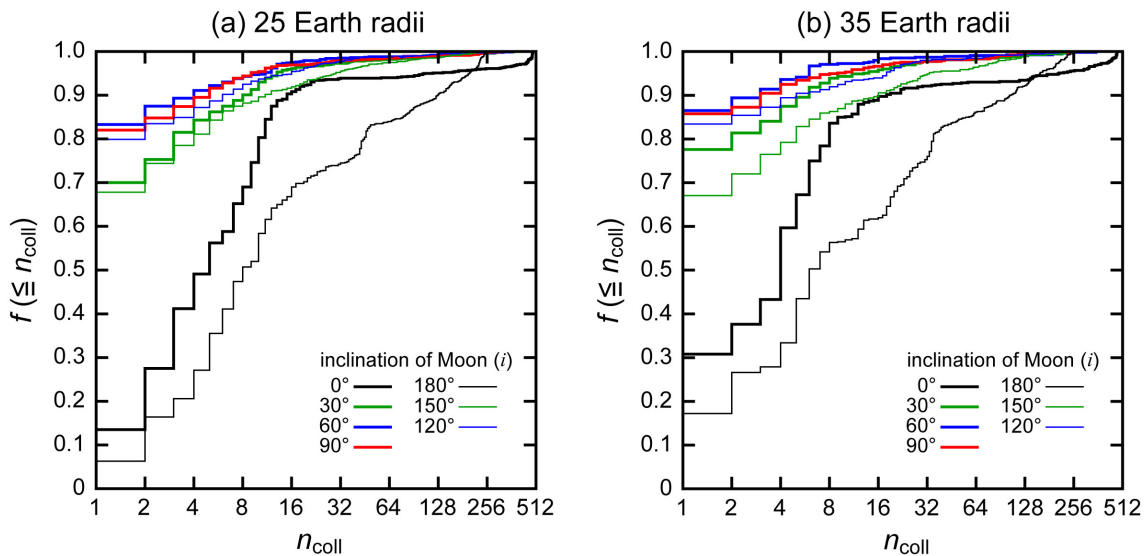

**Figure 4.** Cumulative distribution of impacts with the Moon as a function of inclination and semi-major axis. The Moon is at 25 Earth radii in panel (a), and at 35 Earth radii in (b). The various colors correspond to different lunar inclinations ($i$). Only when the Moon is in the plane of the ejecta ($i = 0°$ and $180°$) are a substantial number of collisions possible.

The results of our analyses are reported in **Figure 4**, which shows the cumulative number of impacts as a function of the initial lunar inclination and the semi-major axis. In the top panel the Moon is at 25 Earth radii while in the bottom panel it is at 35 Earth radii. Results show that only in the case when the Moon's orbital plane is close to that of the ejecta do we register potentially many collisions; for inclined orbits at best a few collisions are scored.

Each SPH particle has a total mass of $10^{-6}$ Earth masses. The Moon's HSE budget requires it to have accreted approximately 0.025 wt% or $3 \times 10^{-6}$ Earth masses, or fewer than 3 SPH particles. From **Figure 4** we see that this is satisfied at least 75% of the time unless the impact occurred in the Moon's orbital plane, at which time the probability is expected to be $1/2 \sin^2 i \, di$, or about 0.25 for $i < 30°$ or $i > 150°$. In summary, there is a reasonably good (~ 20%) chance that the Moon's HSE contribution can be accounted for by sesquinary collisions with debris generated from the impactor that supplied Earth's LV.

## 4 Discussion

Here we assess the fates of the ~10-m iron fragments (**Figure 3**) during re-accretion onto Earth, and explore their potential as the source of HSEs in the post-impact terrestrial mantle. We might expect that ~10-m size iron blobs experience further fragmentation during re-accretion. If iron materials are ejected from the forming crater during re-accretion, the typical size of ejected fragments can be re-estimated from Equation (1). In our calculations, we estimate the strain rate as roughly $\dot{\varepsilon} \sim v_{imp}/D_{imp} = 500$ s$^{-1}$, where we considered $v_{imp} \sim 5$ km/s and $D_{imp} = 10$ m. For the case of 45-degree impact, 73% of iron SPH particles that experienced fragmentation during the first contact of the collision later re-accrete onto the Earth's surface at > 5 km/s (~ half the Earth's escape velocity). On the other hand, if iron core fragments are not ejected, but are stuck in the forming impact melt pool, the size of fragments should be controlled by another rule. Kendall and Melosh (2016) found that the core materials entrained in a melt pool become stretched 10 to 100 times their original diameter as re-accreting iron fragments, which implies that 10-m iron impactors are separated into blobs on a scale 1 m or less. These authors considered a differentiated impactor (with iron inside and rock outside) in their simulations, and found that silicate crusts are ejected but major part of the iron core is welded into the crater. We expect, however, that significant amounts of iron would be ejected when the impactor consists only of iron. Thus, expelled 1-mm fragments are expected to blanket Earth's surface as an ejecta layer of iron hail.

If the early Hadean Earth already had liquid water on its surface (Abe, 1993), the following reaction with 1-mm fragments and water would effectively occur:

$$\text{Fe} + \text{H}_2\text{O} \rightarrow \text{FeO} + \text{H}_2. \qquad (5)$$

According to our simulations, approximately 60 wt% of a lunar-sized impactor's core (~ 0.2 wt% of $M_\oplus$) is available to react with (and consume) about 3 $M_{oce}$ water, where $M_{oce}$ 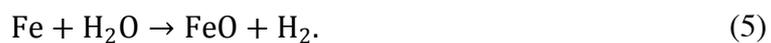 is the present Earth's ocean mass (= $1.4 \times 10^{21}$ kg = 0.023 wt% of $M_\oplus$), in Fe metal oxidation (Eq. 5). Hadean (pre-3.9 Ga) terrestrial zircons have oxygen isotope values consistent with liquid water at Earth's surface in the first ca. 150 Myr (Mojzsis et al., 2001). Planet formation models also support early delivery of water to the growing Earth (O'Brien et al. 2014; Rubie et al., 2015). Thus, it seems reasonable to postulate that oceans existed during the LV phase. If the early Hadean Earth had adequate water in its surface reservoirs (crust and upper mantle), we expect that a huge amount of FeO can be formed. This FeO layer is expected to be mixed into the mantle by any number of processes proposed to have been active on the early Hadean Earth including global resurfacing (Marchi et al. 2014) and/or vigorous crustal recycling driven by higher heat flow (e.g. Rozel et al., 2017).

Barring the availability of 3 $M_{oce}$ on Earth to oxidize all the re-accreted metallic iron in our impact model, partial oxidation could still proceed after metallic iron enters the Earth's mantle. This would occur if the Hadean mantle was already relatively oxidized. A number of studies (Trail et al., 2011; Delano, 2001; Frost et al., 2008) support the notion that the oxidation state in the Hadean Earth's mantle was buffered at fayalite-magnetite-quartz (FMQ). When 1-mm iron fragments enter the Earth's mantle, those would be quickly oxidized. The oxidation timescale can be estimated to be just 1 day from $d^2/D$, where $D$ is the diffusion coefficient of oxygen in Fe oxides (> $10^{11}$ m$^2$/s, e.g., Takada and Adachi 1986). If we presume that all metallic Fe in a lunar-sized impactor is becomes oxidized into FeO in the above process, the amount of FeO is estimated to be $2.3 \times 10^{22}$ kg. This corresponds to ~0.6wt% of FeO in the Earth's mantle, which is well below the FeO content in the current Earth's mantle (~ 8wt%; O'Neill and Palme, 1998).

The reaction between Fe and water also produces $H_2$. Reduction of 3 $M_{oce}$ yields a $pH_2$ of about 90 bars expressed as surface atmospheric pressure. We postulate that the intense extreme UV radiation from the young Sun (Ribas et al., 2010) erodes a 90 bar $H_2$-atmosphere through hydrodynamic escape in about 200 Myr (Genda et al. 2017a; Hamano et al. 2013). Such post-LV collision hydrogen-rich atmosphere from colossal impact and core disruption has obvious implications for biopoesis on Hadean Earth (e.g. Urey 1952; Miller 1953), and Mars.

As with Earth, debate continues over whether an LV occurred on Mars. Walker (2009) summarized evidence for a martian late accretionary augmentation of ~0.7wt%, which is comparable relative to the Earth's; the terrestrial value for an LV complement was recently upwardly revised to be about 0.8 wt % (Day et al., 2016). It is worth noting that Righter et al. (2015) argue that Mars' mantle HSEs can be solely explained by metal-silicate equilibration. In their exploration of the probable nature of a martian LV, Brasser and Mojzsis (2017) explained that the red planet's HSE abundances are explicable by impact of a Ceres-sized object before about 4.43 Ga by which time martian crust formation was well underway (Humayun et al., 2013). Further, Brasser and Mojzsis (2017) suggested that this colossal impactor could explain both the martian hemispherical dichotomy (Andrews-Hanna et al., 2008; Bottke and Andrews-Hanna, 2017) and at the same stroke provides a source for the martian satellites (e.g. Rosenblatt et al. 2016). In addition, Mars appears to have formed from planetary building blocks that were more volatile-rich than that which were supplied to the growing Earth, and probably Venus (e.g. Brasser et al., 2017). Accordingly, we predict that oxidation of metal (Eq. 5) in a martian LV impact with the pre-Noachian hydrous lithosphere or upper mantle also generated an early hydrogen-rich atmosphere.

# 5 Conclusions

We report the output of SPH-based numerical simulations for impacts of a lunar-sized object onto the early Hadean Earth as an explanation of the anomalous enhancement in HSE abundances of Earth's present mantle. The fate of the impactor's core, which should have carried most of the HSE payload of the impactor, depends strongly on impact angle. For a near head-on collision ($\theta \sim 0°$), the impactor's core simply merges with the Earth's core. On the other hand, for an oblique impact, the impactor's core becomes elongated during the post-impact phase and a significant fraction of it is ejected from the impact site. In this process, a large fraction of impactor material experiences fragmentation. The typical size of fragments was estimated to be ~ 10 m. At the statistically most likely impact angle of $\theta = 45°$, the total mass of fragments that are available to re-accrete to Earth is maximized (~ 60 wt% of the impactor's core).

Results show that under a range of impact conditions, most of the fragmented core materials experience a secondary fragmentation and shearing regime prior to falling to Earth in a hail of molten iron. The quantity of iron fragments that accrete onto the Moon as sesquinaries is low, but enough to account for the relative difference in HSE abundances between the two bodies.

In the post-impact environment of a Late Veneer scenario described in this work, 10-meter size iron blobs are further fragmented into 1-mm droplets at re-accretion. We propose that oxidation of these small iron droplets by the early Hadean Earth's surface zone (crust and/or mantle) is a physically and chemically plausible supply mechanism for suspending HSEs in the impactor's core to the Earth's mantle. In our model, a single oblique impact at the statistically most-likely angle ($\theta \sim 45°$) of a lunar-sized object onto Earth can account for the delivery of HSEs to the terrestrial mantle. This would be the case provided that the planet had sufficient water oceans, or the Hadean Earth's mantle was near present oxygen fugacity, or both. We postulate that Pre-Noachian Mars should have experienced similar effects if it witnessed a colossal impact that delivered its HSEs (Brasser and Mojzsis, 2017) which then reacted with an extensive hydrous surface zone.

**Acknowledgments**
We thank Shigeru Ida for his helpful advice on our numerical simulations. Editorial stewardship by Fred Moynier, and constructive comments by two anonymous reviewers improved this work. This study was supported by Grant-in-Aids for Scientific Research from the Japan Society for Promotion of Science (15K13562, 17H02990, 17H06457) to HG and (16K17662) to RB. RB and SJM acknowledge support from the Collaborative for Research in Origins (CRiO) which is supported by The John Templeton Foundation – FfAME Origins program: The opinions expressed in this publication are those of the authors, and do not necessarily reflect the views of the John Templeton Foundation. SJM is also grateful to the Earth-Life Science Institute (ELSI) at the Tokyo Institute of Technology for sabbatical support during which time this project was realized. SJM further acknowledges support from the NASA Exobiology Program (NNH14ZDA001N-EXO) for our on-going investigations of terrestrial-type planetary bombardments. The source codes for the models used in this study are archived at the Earth Life Science Institute of the Tokyo Institute of Technology. The data, input and output files necessary to reproduce the figures are available from the authors upon request.